# Floating zone furnace equipped with a high power laser of 1kW composed of five smart beams

**Yoshio Kaneko [a,*], Yoshinori Tokura [a,b]**

[a] RIKEN Center for Emergent Matter Science (CEMS), Wako 351-0198, Japan
[b] Tokyo College and Department of Applied Physics, University of Tokyo, Tokyo 113-8656, Japan

*Corresponding author.
Address: 2-1 Hirosawa, Wako, Saitama 351-0198, Japan.
E-mail address: yskaneko@riken.jp (Y. Kaneko).

**ABSTRACT**

A floating zone furnace is a powerful tool for single crystal growth. Here, we report the floating zone furnace using a high power diode laser of 1kW composed of five 200 Watt heating beams. The five beams with the designed irradiation intensity distribution enable the melt-growth of single crystals for refractory materials having a high melting temperature above 2000°C, volatile materials and even incongruent materials. The power of five beams in the horizontal plane provides uniform intensity in the outer circumference of the raw material rods, while the vertical direction power of five beams can be made to have a bell shape irradiation intensity so as to relax residual thermal stress in the grown single crystal. The present furnace enables one to directly monitor the temperature of the hot spot of raw material and the molten zone within a 1 mm diameter. Such an in-situ monitoring function of the temperature on the molten zone greatly enhances the controllability for optimizing the crystal growth condition.

# 1. Introduction

The floating zone (FZ) method using halogen lamps [1,2] or a xenon lamp [3,4] has been a powerful tool for the researchers in the field of the strongly correlated electron systems to grow single crystals of transition-metal oxides. However, the conventional FZ method is difficult to apply to refractory materials with a high melting temperature, volatile materials or incongruent materials. To partly overcome this difficulty, the FZ method using a diode laser, referred to as the LFZ method hereafter, has recently been proposed [5]. This LFZ instrument, composed of seven diode lasers with each power of 50W, has successfully shown many advantageous properties over the conventional halogen lamp or xenon lamp FZ method. Nevertheless, the total power with 350W is still insufficient, and development of more high power of LFZ with optimal light intensity distribution and monitoring system of temperature in a molten zone may lead to the successful crystal growth, which would be otherwise difficult. Here we describe the design of a new LFZ furnace with a higher total power of 1kW consisting five beams with the designed irradiation intensity distribution as well as its rich accessories to enable many improved achievements and show the successful examples of the growth of exotic single crystals in terms of the high-power LFZ method.

# 2. Setup of high-power laser floating zone furnace

In the presently developed LFZ furnace, the five beams at a wavelength of 940nm with each power of 200W are guided via an optical fiber from a diode laser (LD) source. The total laser beam intensity is designed at 1 kW to grow single crystals of refractory materials having a high melting temperature exceeding 2000°C. In the LFZ, a raw material rod is heated by using a laser beam emitted from a beam-head composed of several lenses to realize the optimal irradiation light. The LFZ must have a water-cooled beam-damper that receives the high-power laser beam that has passed around the material rod in the propagating direction of the laser beam. The conventional LFZ equipment uses seven laser beams [5, 6]. In the case of using seven beams, seven beam-heads and seven beam-dampers must be arranged on a single optical bench in LFZ. The present LFZ as developed here is composed of five beam-heads and five beam-dampers so as to simplify the optical system. To compensate the reduced number of laser beams, we implemented a LD with a high power of 200W. The simplified optical system makes it easy to install a radiation thermometer and other beneficial accessories, as described later. As we will see in the next section, with the design change from seven beams to five beams there occurs no problem concerning the horizontal uniformity.

Figure 1 (a) shows the experimental setup of the developed LFZ and the magnified view around the sample rod. Figure 1 (b) is the photograph around the sample. Five beam-heads make five laser beams into the desired irradiation intensity distribution shape. It is easier with the LFZ to design the irradiation intensity distribution of the heating laser beam than the halogen lamp. This is a great merit of using the laser beam as a heating light source for crystal growth.

Two types of laser beam profiles, referred to as type A and type B, were prepared. The irradiation intensity distribution of the heating beam on the two-dimensional plane (X, Y) perpendicular to the Z-axis direction of beam propagation is precisely created by using each laser beam-head consisting of 10 lens sets designed by using the optical simulator. The Y-axis direction is the vertical direction as well as the center axis direction of the sample rod (see Fig. 1 (a) ). The X-direction is the horizontal direction as well as the azimuthal direction of the raw material rod (see Fig. 1 (a) in the case of No.3 beam-head). Each irradiation intensity distribution on the two-dimensional plane (X, Y) is the completely same for the five laser beams. The irradiation patterns, type A and type B, on the two-dimensional plane (X, Y) of one laser beam is shown in Fig. 2 (a) and Fig. 2 (b), respectively. Laser beam irradiation intensity (I) with the vertical axis in this figure is plotted in X-Y two-dimensional coordinates. The irradiation intensity (I) has been measured by using a CCD camera with which has an effective area of 10mm*12.5mm with the pixels of 4.54μm*4.54μm (LT665-1550, MKS Instruments Inc.). By measuring the radiation intensity shapes of the five beams having different incident directions with a CCD camera placed on the central axis of the sample rod, it is possible to confirm that the five beam intensity distributions are exactly the same. These two kinds of irradiation intensity distributions can be easily interchanged only by adding one cylindrical lens to each beam-head. For both cases of type A and type B, the irradiation intensity distribution along the X-axis direction is optically designed to be uniform to ensure high uniformity of heating in the azimuthal direction of the raw material rod. Only the irradiation intensity distribution in the Y-axis direction is different; i.e. a flat shape for type A or a Gaussian shape for type B, as described in detail below.

First, let us show the design of the irradiation intensity in the X-axis direction. The beam length in the horizontal (X) direction is required to be larger than the diameter of the crystal rod to be melted. It is typically the width of ±4mm in the X-direction of both type A with a flat shape and type B with a gaussian shape. Figure 3 shows the comparison of the heating beam intensity in the azimuthal direction of the raw material rod between the cases of the present LFZ furnace with five laser beams of type A with a flat shape or type B with a gaussian shape and the halogen lamp FZ furnace with a confocal elliptical mirror equipped with four halogen lamps. With a confocal elliptical mirror, even in the case of using four halogen lamps, the uniformity remains about 75%. The green dashed line in Fig. 3 is for the case of the halogen lamp FZ with a confocal ellipsoidal mirror with four halogen lamps, as calculated by the following procedure: First, the irradiation intensity Ha on a sample rod from a halogen lamp placed at one focal point of a confocal ellipsoid is measured by using a plate-like thermocouple placed at the center position of a sample rod. The irradiation intensity distribution Ha(Y) on the Y-axis is measured by moving the thermocouple on the Y-axis direction. The obtained Ha(Y) on the Y-axis has a gaussian shape. The position at 10% of the maximum intensity of the halogen lamp in the Y direction is ±25mm from the center at Y-axis. The Ha(X) along the X-axis is the same as the Ha(Y) along the Y-axis. The green dashed line is calculated by assuming the four irradiation light beams with the same Ha(X). With a confocal elliptical mirror in the case of using two halogen lamps, the uniformity will be even worse, less than 50%. On the other hand, the red solid line in Fig. 3, which was calculated and reported also by Ito et al.[5], shows irradiation intensity distribution in the azimuthal direction of the raw material rod in both cases of type A with a flat shape and type B with a gaussian shape in the case of using LFZ. In this case, high uniformity of 95% or more can be ensured over 360° around the raw material rod. For the actual crystal growth procedure in the FZ method, the crystal rod is rotated. As a result, in the Ha-FZ case, large temperature fluctuations near the interface between the crystal-growing region and the molten liquid phase becomes serious. However, the high uniformity of the irradiation intensity of the heating light from the azimuthal direction of the crystal rod in both cases of type A with a flat shape and type B with a gaussian shape can dramatically suppress such temperature fluctuation which would occur in the Ha-FZ. The conventional LFZ equipment uses seven laser beams as a heating beam for raw material rod. The uniformity of 97% or more can be ensured over 360° around the raw material rod for this case. The change from seven beams to five beams leads to the slight change of the calculated uniformity to 95%, however, this minor reduction in the uniformity is turned out to cause no problem in sample examples of actual crystal growth.

The next issue is the difference between type A with a flat shape and type B with a gaussian shape in the irradiation intensity along the Y-axis direction. The beam of type A with a flat shape (Fig. 2 (a)) has uniform intensity distribution in a region within the height of ±2mm in the Y direction. For the irradiation intensity distribution of type B with a gaussian shape (Fig. 2 (b)), on the other hand, the lens system of the irradiation head is designed so that the irradiation intensity distribution in the Y-axis direction of the laser beam may have a gaussian form, which has a maximum intensity $I_{max}$ at the center of the vertical molten zone. The region where the intensity exceeds 10% of $I_{max}$ is set within ± 5mm, ± 10mm, or ± 20mm in the vertical direction from the center of the molten zone. The cylindrical lenses with different focal lengths are designed so that the irradiation intensity along the Y-axis direction has a gaussian form on the central axis of the rod. The ± 20mm gaussian form is almost the same as the case of Ha-lamp with ± 25mm gaussian form. Such a special design of the beam intensity distribution profile as type B with a gaussian shape is unique in the present LFZ system and plays a key role in the success of crystal growth which would be difficult with the conventional LFZ method with type A beam with ±2mm flat shape beam. These three kinds of B type with a gaussian shape can be easily interchanged only by adding one cylindrical lens to each beam-head for type A beam with ±2mm flat shape. We can easily select the heating beam profile with optimal irradiation intensity distribution depending on crystal-growth materials.

To see the effect of the beam pattern shaping, we show the case of LFZ crystal growth of the perovskite Mn oxide crystals done by using the beam profiles of type A with ±2mm flat shape and type B with ±10mm gaussian shape along the vertical direction. For crystal growth of TbMnO$_3$ as a typical example of the perovskite Mn oxide, we adopt the following growth conditions; 1 atmosphere of argon gas and a growth rate of 2mm/h. Figure 4 (a) shows cross-sectional pictures of TbMnO$_3$ single crystals. A cross-section of the single crystal LFZ-grown with type A with ±2mm flat shape shows many cracks. The irradiation intensity distribution of type A with ±2mm flat shape as used in the conventional LFZ [5] has a steep temperature change at the boundary between the molten part and the crystalized part. This abrupt temperature variation causes the residual thermal strains in the grown single crystal. The residual strain sometimes produces cracks in the grown single crystal and causes serious problems, e.g. impairing the reproducibility of measured physical properties, such as charge transport characteristics. On the other hand, the irradiation intensity distribution of type B with the gaussian irradiation shape has a gentle temperature change across the boundary between the molten and the crystallized parts. Such a gradual temperature distribution can dramatically suppress the residual thermal strains in the grown single crystal. As a result, the cross-section of the single crystal made by using the type B with ±10mm gaussian shape shows only a few small cracks. This is an important performance for fabricating a single crystal of an insulator with a low heat conduction coefficient.

One of the advantages of using the present LFZ furnace is that the molten zone temperature can be monitored during crystal growth. Unlike the conventional halogen-lamp FZ or xenon-lamp FZ apparatus, which requires a confocal ellipsoid mirror covering the upper and lower parts of the rod, the LFZ method with five beams has the merit that it is possible to secure a space where a radiation thermometer can be easily installed to directly monitor the temperature within the diameter of 1 mm in the molten zone in real-time during the crystal growth. We used a radiation thermometer with an InGaAs sensor. The InGaAs sensor is sensitive to a wavelength of 1.55μm while being insensitive to a wavelength of 940nm of the laser. Thus, the influence of the laser beams scattered from the surface of the quartz tube and/or sample can be eliminated. The measurable temperature range is from 300°C to 3500°C, and the reproducibility of the measured temperature is within ±0.5°C (IR-CZQH7N, Chino Co.).

Figure 4 (b) shows the monitor image (left photograph) of the molten part using of M-type hexaferrite $BaFe_{12}O_{19}$ during the LFZ crystal growth. We adopt the following growth conditions; 10 atmospheres of oxygen gas, a growth rate of 2mm/h and rotation speed with ±20 rpm of the upper and lower rods. $BaFe_{12}O_{19}$ shows congruent melting under 10 atmospheres of oxygen gas atmosphere. In the present case is used type B with ±10mm gaussian shape along the vertical direction. Figure 4 (b) is a photograph taken to observe the temperature distribution when the upper and lower rods are stationary without rotating. The color change in the monitor image of the molten zone on the left photograph is smoothly darkened vertically from the brightest part in the center. It is indicating that the temperature distribution in the vertical direction is changing smoothly. The horizontal brightness in the azimuthal direction of the molten zone, the upper rod and the lower rod is uniform. It shows that the temperature in the azimuthal direction of the molten zone, the upper rod and the lower rod is uniform.

The area enclosed by the small red open circle within the molten zone on the left photograph indicates the area whose temperature is monitored in situ by the radiation thermometer. The trend chart (right panel) exemplifies the temperature of the local spot (1mmφ) measured within the molten zone in real-time during the crystal growth of ferrite materials. The melting temperature of $BaFe_{12}O_{19}$ is 1535°C under 10 atmospheres of oxygen gas atmosphere. The laser intensity is adjusted so as to determine the optimum temperature for forming the molten zone in a few hours at the initial run. After that, the molten zone temperature shows remarkable stability over several hours, which is maintained within ± 2°C, even when the upper feed rod and lower single-crystal rod are counter-rotating at 20rpm to each other.

The direct monitoring of the molten zone temperature can remarkably advance the floating zone crystal growth method. It is also possible to directly observe the temporal reduction in the temperature with high accuracy due to the accretion on the inside of the quartz tube, therefore the laser power can be rapidly fed back to maintain the optimum temperature of the molten zone before unstable of the upper rod and/or molten zone. If the phase diagram of the material is known, it is possible to quickly set the laser intensity for obtaining an optimum temperature of the molten zone for growing the crystal of the material.

## 3. Examples of single crystal growth with high-power laser floating zone furnace

### 3.1 Refractory materials

#### 3.1.1 Ruby:

A photograph of ruby ($Al_2O_3$ doped with 3% Cr with melting temperature 2045°C) single crystal made by using the present LFZ furnace is shown in Fig.5 (a) as a typical example of refractory oxide compounds with high melting temperature. For growing these crystals of Ruby, we adopt the following growth conditions; 1 atmosphere of air and a growth rate of 7mm/h. This case is used type B with ±10mm gaussian shape along the vertical direction. In type A beam with ±2mm flat shape case, a few cracks often occur. The obtained crystal is a single crystal with a rod diameter of 8mm and a rod length exceeding 70mm. It grows into a quadrangular prism with four well-defined crystal planes. With the use of a halogen-lamp FZ furnace, however, such a clean crystal face can rarely be obtained. The surface of single crystals grown by the LFZ method frequently shows such shiny crystal facets. This is because the irradiation intensity distribution of the laser beam in the azimuthal direction of the molten zone shows high uniformity and hence the temperature fluctuation at the interface between the solid phase and the liquid phase during the rotation of the crystal rod is minimized. The uniformity is much superior to the case of the halogen-lamp FZ furnace which suffers from a relatively large temperature fluctuation around the interface region due to the rotation of the crystal rod and forms a roughened crystal surface. In terms of the present LFZ, we can avoid the unevenness composed of multi-crystal domains which would adversely affect the crystal growth of the crystal rod.

#### 3.1.2 $SmB_6$:

$SmB_6$, which has a high melting point (2350°C), has recently been attracting revived interest as the correlated topological insulator [7]. We have tried to grow a high-quality crystal with such a high melting point by using the present LFZ furnace. For growing these crystals of $SmB_6$, we adopt the following growth conditions; 1 atmosphere of argon gas with a gas flow rate of 2000cc/m and a growth rate of 20mm/h. In this case, LFZ is using type B with ±5mm gaussian shape along the vertical direction. $SmB_6$ has a high melting point and high thermal conductivity. The LFZ power needs a higher optical density of the beam. Therefore, a heating beam with a narrow distribution along the vertical direction is required. This LFZ is also a powerful tool for growing crystals of materials other than oxide compounds. This LFZ has a high vacuum system with a turbo molecular pump and can flow high purity argon gas after getting a high vacuum level below $1*10^{-3}$ Pascal in the quartz tube. The large flow rate of high purity argon gas was increased to prevent the evaporated material from adhering to the inner surface of the quartz tube. The obtained single crystal has a shiny mirror-like surface as shown in Fig.5 (b) (left panel). The Laue pattern on the crystal surface (Fig. 5 (b), right panel) shows that the crystal facet surface is (111) plane showing three-fold symmetry. The reason for such a successful growth is the same as in the above case of ruby crystal. $SmB_6$ single crystal made by using FZ furnace with a confocal ellipsoidal mirror with four Xenon lamps [8] does not show such a shiny mirror-like facet surface as grown by this LFZ furnace.

## 3.2   Incongruent melting materials
### 3.2.1 Y-type hexaferrite:

Y-type hexaferrite single crystals have been conventionally grown by a flux method [9] because these materials show incongruent melting. Nevertheless, we have succeeded in growing a large single crystal [10-12] of Y-type hexaferrite $Ba_2Co_2Fe_{12}O_{22}$, known as a multiferroic compound of spin origin, by using the present LFZ. We adopt the following growth conditions; 10 atmospheres of oxygen gas atmosphere and a growth rate of 1mm/h. For the crystal growth, a seed crystal having a desired composition and crystal structure is required. With this, the desired crystal composition is epitaxially grown on the seed crystal, when a stable temperature is kept along the Y-axis direction around the solid(crystal)-liquid(molten) interface. A-type with ±2mm flat shape has been used with good enough results; there are only one or two and smaller cracks in the grown rod, even if they occur. The Y-type hexaferrite $Ba_2Co_2Fe_{12}O_{22}$ may be resistant to thermal strains in a crystal growing, unlike the perovskite Mn oxide crystals. To increase the resistivity of the samples, the cut pieces of obtained crystal were annealed in 10 atmospheres of oxygen gas atmosphere at 1000°C for 100 h in a sealed quartz tube.   We also tried to prepare a single crystal of Y-type hexaferrite $Ba_2Co_2Fe_{12}O_{22}$ by using the Ha-FZ with two halogen lamps. The seed rod with a Y-type crystal structure without other crystal structures can be obtained after growing the rod in several centimeters in the case of Ha-FZ. However, many domains with different crystal-axes are formed in the seed rod. This is because the uniformity of temperature at the periphery of the seed rod is poor and the crystal domains with different crystal orientations always grow from the periphery of the seed rod, in contrast with the case of the LFZ method. The present LFZ may have more possibility of growing single crystals with incongruent melting in terms of the traveling solvent floating zone method. For example, the successful applications of the LFZ method to the growth of copper oxide compounds relating to high-temperature superconductors are anticipated, as already demonstrated for one of the layered copper oxychlorides [13].

## 3.3 Highly volatile materials
### 3.3.1 $Nd_2Mo_2O_7$ :

Pyrochlore-type $Nd_2Mo_2O_7$ (NMO) shows one of the interesting materials which show the large topological Hall effect arising from the scalar spin chirality characteristic of pyrochlore lattice [14]. For growing these crystals of NMO, we adopt the following growth conditions; 1 atmosphere of argon gas and a growth rate of 2mm/h. In growing NMO, type A with ±2mm flat shape has been used with good enough results; there are only one or two small cracks in the grown rod, even if they occur. NMO may be resistant to thermal strains in a crystal growing. A single crystal of NMO can be produced with the growth temperature above the melting temperature of 1630°C.　For the crystal growth, a seed crystal having a desired composition and crystal structure is required. This compound shows high volatility near the melting temperature. The irradiation intensity of the heating beam is partly interrupted by the deposit (mainly $MoO_2$) from the molten zone to the inner surface of the quartz tube within growing the crystal of NMO. As a result, the molten zone temperature may be lowered. We can quickly detect the temporal reduction in the temperature with high accuracy due to the accretion on the inside of the quartz tube before unstable of the upper rod and/or molten zone. And then, we can rapidly control to increase the temperature of the molten zone by adjusting the laser power. Figure 6 (c) shows a picture of a quartz tube after the crystal growth. The blue deposit on the quartz tube wall is $MoO_2$. However, the tube wall portion irradiated by the laser beam appears transparent.　The evaporated materials adhering to the inner surface of the quartz tube is reheated by the laser light and evaporates again to ensure the transparency of the inner surface of the quartz tube. $MoO_2$ begins to evaporate at around 1100°C. The reason why this re-evaporation phenomenon with $MoO_2$ occurs is that the laser beam having an optical density with several orders of magnitude higher than that of the Ha lamp can rapidly heat to above 1100°C the $MoO_2$ adhering to the transparent quartz tube. After 2 to 3 hours, the rate of the deposit to the inner surface of the quartz tube and the re-evaporation of the deposit on the inner surface of the quartz tube becomes constant. The transmittance of the quartz tube of the beam is constant, and the temperature of the molten zone is constant without increasing the beam intensity. The quartz tube is not completely transparent, and a small amount of deposit material is attached. Stable crystal growth has been realized without the fluctuations of the material rod or the molten zone. Thus, the NMO crystal can be steady grown for long hours without increasing the heating beam intensity. This phenomenon shows that the LFZ method has a great advantage in obtaining single crystals of highly volatile oxide materials.

### 3.3.2 $SrRuO_3$:

$SrRuO_3$ (SRO) is an archetypal perovskite ferromagnetic metal of great interest [15,16]. For growing the crystals of SRO, we adopt the following growth conditions; 10 atmospheres of argon gas with 10% oxygen gas and a growth rate of 7mm/h. This case is used type B with ±10mm gaussian shape along the vertical direction. SRO shows high volatility and is decomposed in the molten state. $SrRu_{1.05}O_3$ in excess of Ru was used as the raw material rod. Therefore, the irradiation intensity of the heating beam is partly interrupted by the deposit (mainly $RuO_2$) from the molten zone to the inner surface of the quartz tube. As a result, the molten zone temperature may be lowered. And then, we can rapidly control to increase the temperature of the molten zone by adjusting the laser power before the molten zone is unstable. This is the same as that of NMO crystal growth. In this case, $RuO_2$ adhering to the inner surface of the quartz tube is re-evaporated by laser beam, by which the transparency of the inner surface can be secured to constant. $RuO_2$ begins to evaporate at around 1000°C. A laser beam with a high optical-density can rapidly heat $RuO_2$ adhering to a transparent quartz tube to over 1000°C. Thus, the SRO crystal can be steadily grown without increasing the beam intensity after the starting process for 1~2 hours in this LFZ method. For getting a single crystal of SRO, the molten zone temperature of the $SrRu_{1.05}O_3$ feed rod has been needed to set at around 1600°C. When melting $SrRu_{1.05}O_3$ feed rod at a temperature of 1700°C, $Sr_3Ru_2O_7$ crystal has been obtained. When melting $SrRu_{1.05}O_3$ feed rod at a temperature of 1800°C or higher, $Sr_2RuO_4$ crystal has been obtained as the major crystal phase. Thus, the control of the molten zone temperature plays a decisive role in obtaining SRO single crystals. The deposition of Ru compounds on the inner surface of the quartz tube in the Ha-FZ often causes degradation of the quartz tube, and the removal of the deposit is not easy. However, in the LFZ method, the adhesion of the raw material to the inner surface of the quartz tube is drastically reduced, and the removal of the deposit is very easy.

Finally, we show other examples, La-doped $BaTiO_3$ [17], $MnWO_4$ [18] and CuO, of oxide single crystals obtained by using this LFZ instrument. This LFZ is also a powerful tool for growing crystals of materials other than oxide compounds. This LFZ is equipped with a high vacuum system equipped with a turbo molecular pump. After a high vacuum level is reached in the quartz tube, this LFZ can flow high purity argon gas. Besides $SmB_6$ mentioned above, we have succeeded to make a single crystal of chalcogen materials such as EuS and (Pb,Sn)Te or other material such as RuGe.

# 4. Conclusions

The 1kW high power laser floating zone (LFZ) furnace developed by RIKEN has the following features.

1. The high power 1kW LFZ equipped with five beams with a 200W laser beam shown in this LFZ has been realized.
2. The irradiation intensity distribution of five beams in this LFZ has been smart-designed for the crystal growing. The irradiation intensity distribution of five beams in the horizontal plane provides uniform intensity distribution on the periphery of the raw material rods. The vertical irradiation intensity distribution of five beams can be designed to have a flat shape that can be used with some materials such as Ferrite oxide or Pyrochlore-type oxide, and a bell shape irradiation intensity for relaxation of residual thermal strain in the grown single crystal.
3. The four types of the irradiation intensity distribution of five beams in this LFZ have been available for the crystal growing. One is typed A with ± 2 mm flat shape in the vertical irradiation intensity distribution. The other is three type B with ± 5mm, ± 10mm or ± 20mm gaussian form.
4. The best irradiation intensity distribution of five beams in this LFZ can be selected for growing the crystal of each material and can be easily interchanged by adding one cylindrical lens to each beam-head.
5. Refractory materials with melting temperatures above 2000°C and high vapor pressure can easily be melted by using this high power 1kW LFZ.
6. The single crystal with incongruent melting materials can be grown by using this LFZ.
7. In the case of growing the highly evaporable oxide compound crystals in this LFZ, it was found that the laser beam re-evaporates the oxidized material adhering to the inner surface of the quartz tube and prevents the attenuation of the beam intensity transmitted through the quartz tube.
8. This LFZ, which is equipped with a high vacuum system and can flow high purity argon gas, has succeeded to make single crystals of materials other than oxide compounds such as EuS, (Pb,Sn)Te, $SmB_6$ and RuGe.
9. The melting temperature is *in-situ* measured with pinpoint accuracy within the spot radius of 1mm within the molten zone during the crystal growth process in this LFZ. The measurable temperature range in this LFZ is from 300°C to 3500°C, and the reproducibility of the measured temperature is within ±0.5 °C. This allows for precise temperature control of the molten zone during crystal growth in this LFZ.

The high-power LFZ having the above features can thus provide a powerful tool for the growth of single crystals, which would otherwise be difficult. In the LFZ, the heating part by laser irradiation is a limited area. The desorption of oxygen and water within the LFZ during laser heating may be minimized. Therefore, there is a possibility that an LFZ suitable for single crystal growth of metal alloys, which needs ultra-high pure argon gas atmosphere with oxygen-free, can be provided. In addition, it is possible to provide a high power LFZ having high controllability and capable of measuring the melting temperature for the crystal of metal alloys having a high melting temperature.

**Acknowledgments**

This research was supported by the Japan Society for the Promotion of Science (JSPS) through the "Funding Program for World-Leading Innovative R&D on Science and Technology (FIRST Program)," initiated by the Council for Science and Technology Policy (CSTP). Special thanks to Professor Shinichi Ito, Institute of Materials Structure Science, High Energy Accelerator Research Organization, Japan, for providing the photograph of single crystals of $Nd_2Mo_2O_7$ for neutron scattering experiments.

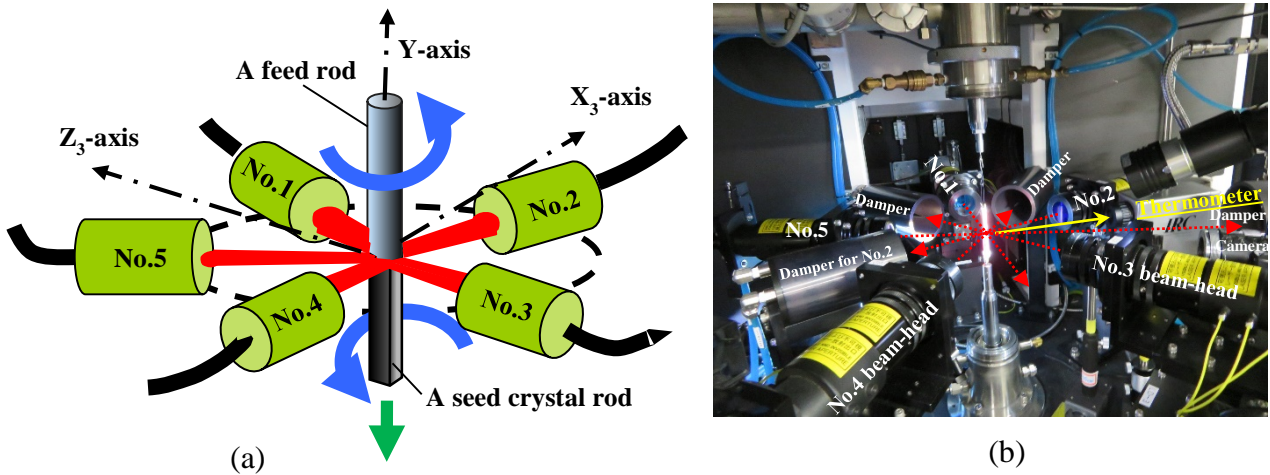

**Fig. 1.** In the figure of (**a**), a feed rod of raw material with a seed crystal rod on the central axis (Y-axis) of the vertical direction is heated from the surroundings with five laser beams exactly arranged at the relative angle of 72° in the horizontal plane. As shown by blue arrows, a feed rod and a seed crystal rod are rotated in opposite directions, while vertically moved (a green arrow) with the controlled gas atmosphere within a cylindrical quartz tube. The five laser beams ensure the uniformity of the outer peripheral temperature of the sample rod. The horizontal $X_3$-axis and the beam propagation $Z_3$-axis for the beam-head of No. 3 are exemplified in the figure. The photograph (**b**) shows the five beam-heads for the five laser beams, four water-cooled beam-dampers for receiving high power laser beams, a camera for observing a molten zone, a feed rod, a seed crystal rod, and a radiation thermometer labeled as "IRC" for directly monitoring the temperature at the pinpoint of the molten zone. The beam-damper for beam-head No.1 placed in front is not shown in this photograph to make the upper rod and under crystal rod to see. This beam-damper can be easily removed when removing the quartz tube. The red dotted arrow overwritten on the photograph shows the invisible laser beam. The beam emitted from the beam-head passes through the sample rod and enters into the water-cooled beam-damper. The yellow solid arrow overwritten on the photograph shows the radiant light emitted from the molten zone and is incident on a thermometer.

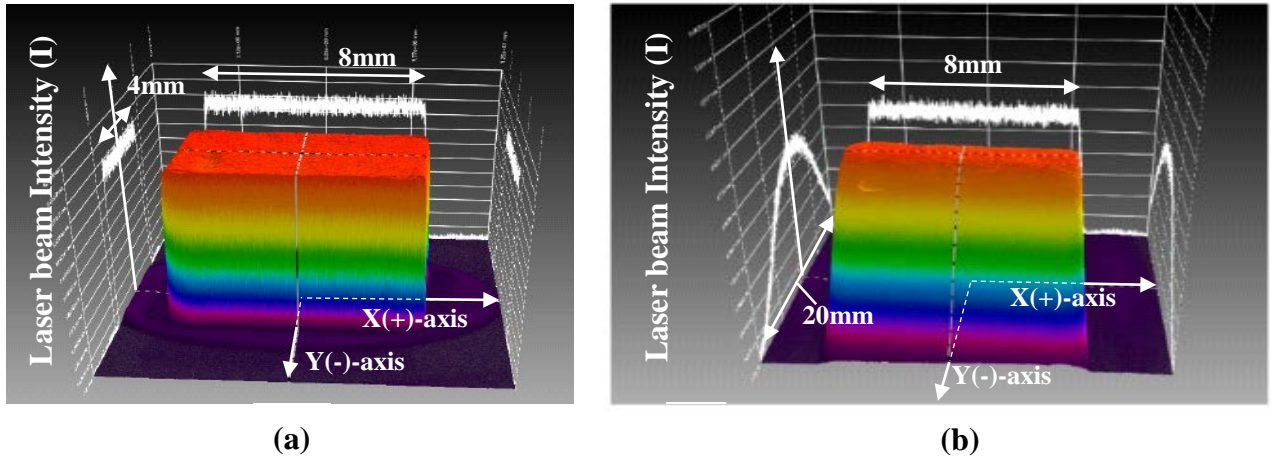

**(a)**                                  **(b)**

**Fig. 2 (a)** It shows the irradiation intensity distribution of type A beam with ±2mm flat shape on a two-dimensional X-Y plane of the irradiation beam perpendicular to the traveling direction of the beam. Laser beam irradiation intensity (I) with the vertical axis in this figure is plotted in X-Y two-dimensional coordinates. The intensity distribution of type A beam with ±2mm flat shape has uniform within the X-Y plane to perpendicular to the propagating direction Z-axis of the laser beam. The length in azimuthal (X-axis) direction of the sample rod is 8mm and the right and left edges position at ±4mm along X-axis. The length in the direction of the central axis (Y-axis) direction of the sample rod is 4mm and the upper and lower edge are ±2mm along Y-axis. The intensity of the laser beam of type A with ±2mm flat shape is designed to abruptly change from maximum intensity to zero intensity at the upper/lower edge of Y= ±2 within a width of about 0.2mm. **(b)** It shows the irradiation intensity distribution of type B with ±10mm gaussian shape on a two-dimensional X-Y plane of the irradiation beam perpendicular to the traveling direction of the beam. The irradiation intensity distribution along the azimuthal (X-axis) direction of the raw material rod direction is optically designed to ensure high uniformity. Type B has gaussian shape intensity along only the Y-axis direction which is the vertical direction as well as the center axis direction of the sample rod. This type B has that the position at 10% intensity of the maximum laser beam in the Y direction is ±10mm from the center at Y-axis. The length which has uniform intensity in the azimuthal (X-axis) direction is kept 8mm. The beam profiles are of the completely same shape for all the five beams.

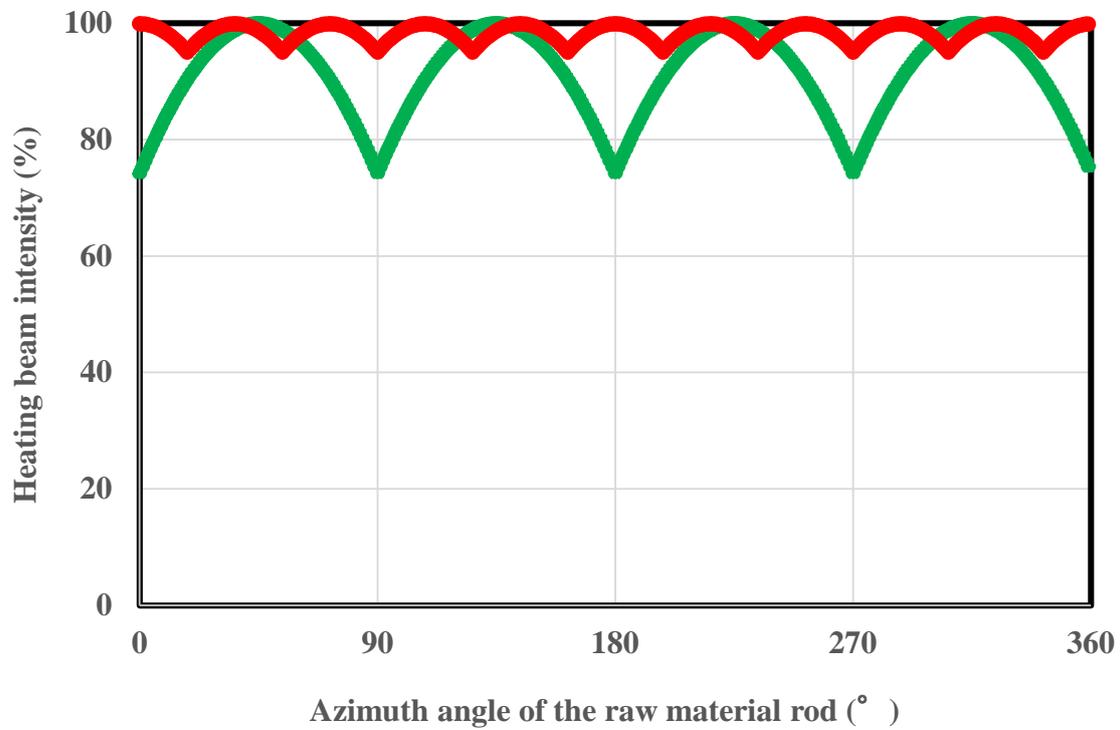

**Fig.3.** Dependence of heating beam intensity on the azimuth angle of the raw material rod. The red solid line [5] shows the case of five laser beams with uniform irradiation intensity distribution in the azimuthal (X-axis) direction in both cases of type A beam with ±2mm flat shape and type B with gaussian shape. The green dashed line is for the case of using a Ha-FZ with a confocal ellipsoidal mirror with four halogen lamps.

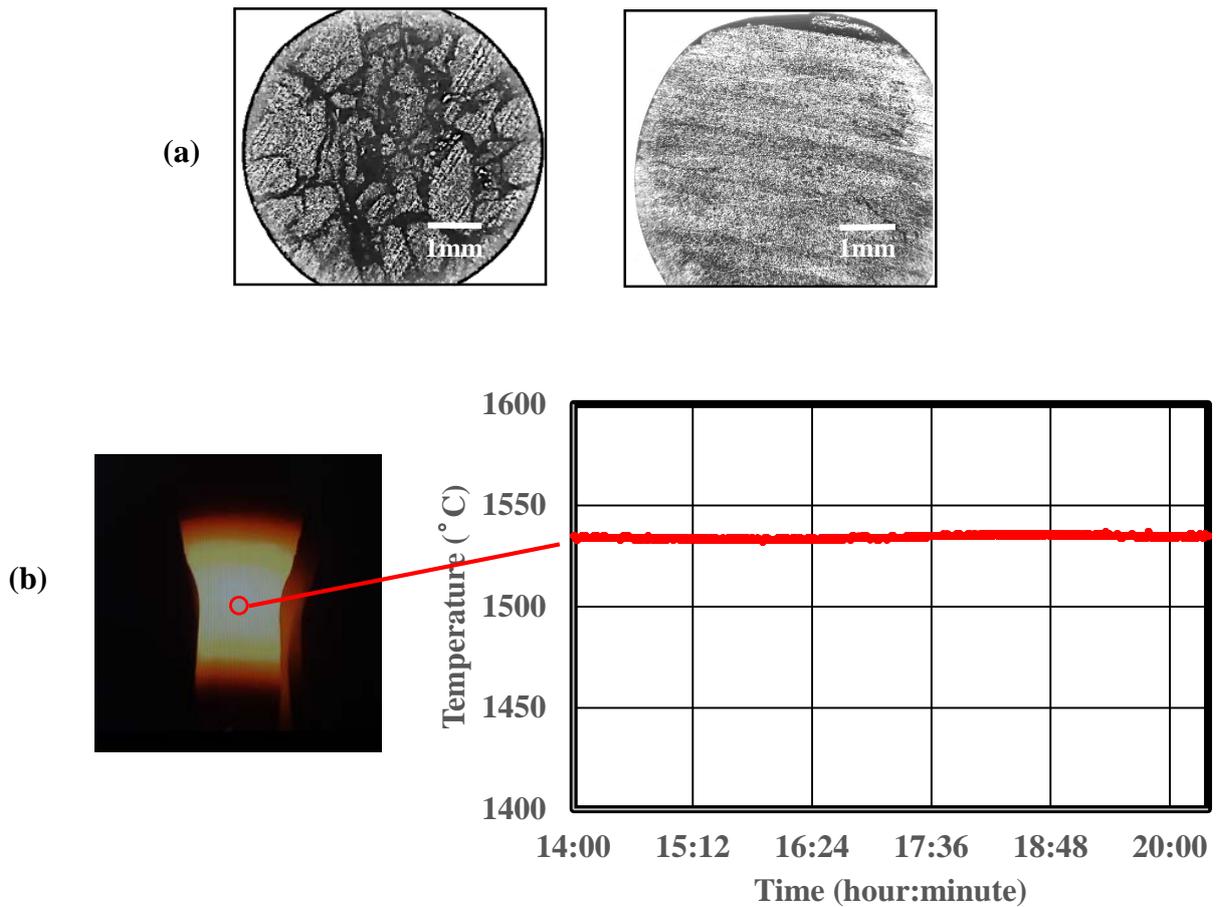

**Fig. 4. (a)** Cross-sectional pictures for TbMnO$_3$ crystals with the use of type A beam with ±2mm flat shape (flat shape along the azimuthal direction (X) and flat shape along the vertical (Y), left) and type B with ±10mm gaussian (flat shape along the azimuthal direction (X) and gaussian shape along the vertical (Y) direction, right). In the case of the use of type A beam with ±2mm flat shape along the vertical (Y) direction and ±4mm flat shape the azimuthal direction (X), there are many cracks in the cross-section (left photograph). In the case of the use of type B with ±10mm gaussian shape along the vertical (Y) direction and ±4mm flat shape the azimuthal direction (X), by contrast, there are only a few small cracks in the cross-section (right photograph). For both cases, the growth rate is 2mm /hr. **(b)** The monitor image (left photograph) of the molten part using of M-type hexaferrite BaFe$_{12}$O$_{19}$ during the LFZ crystal growth. This case is using type B with ±10mm gaussian shape along the vertical (Y) direction. The area enclosed by the small red open circle within the molten zone on the left photograph indicates the area whose temperature is monitored in situ by the radiation thermometer. The trend chart (right panel) shows the temperature of the local spot (1mmφ) within the molten zone measured in real-time.

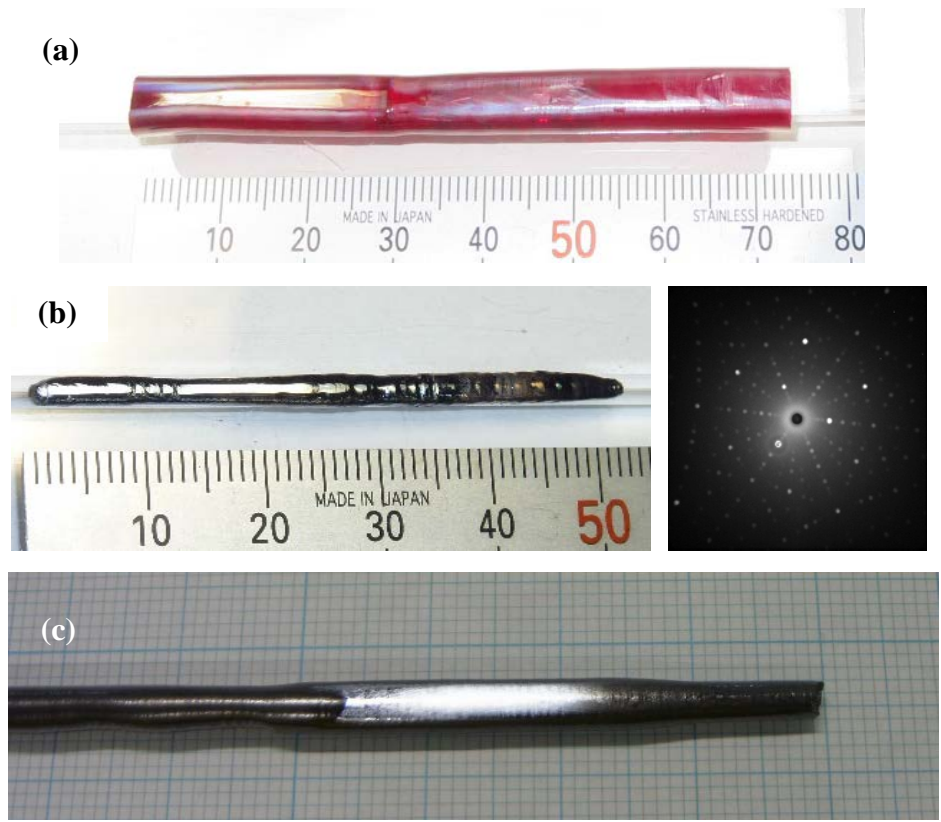

**Fig. 5. (a)** The photograph of a single crystal of ruby ($Al_2O_3$ doped with 3% Cr) with shiny facets. In this case, LFZ is using type B with ±10mm gaussian shape along the vertical direction. **(b)** The photograph of a single crystal of $SmB_6$ with a shiny mirror-like facet. The right panel shows the back Laue reflection from (111) surface of the crystal. In this case, LFZ is using type B with ±5mm gaussian shape along the vertical direction. **(c)** The photograph of the LFZ grown a single crystal of Y-type hexaferrite $Ba_2Co_2Fe_{12}O_{22}$ with a shiny facet. In this case, LFZ is using type A with ±2mm flat shape in the vertical direction.

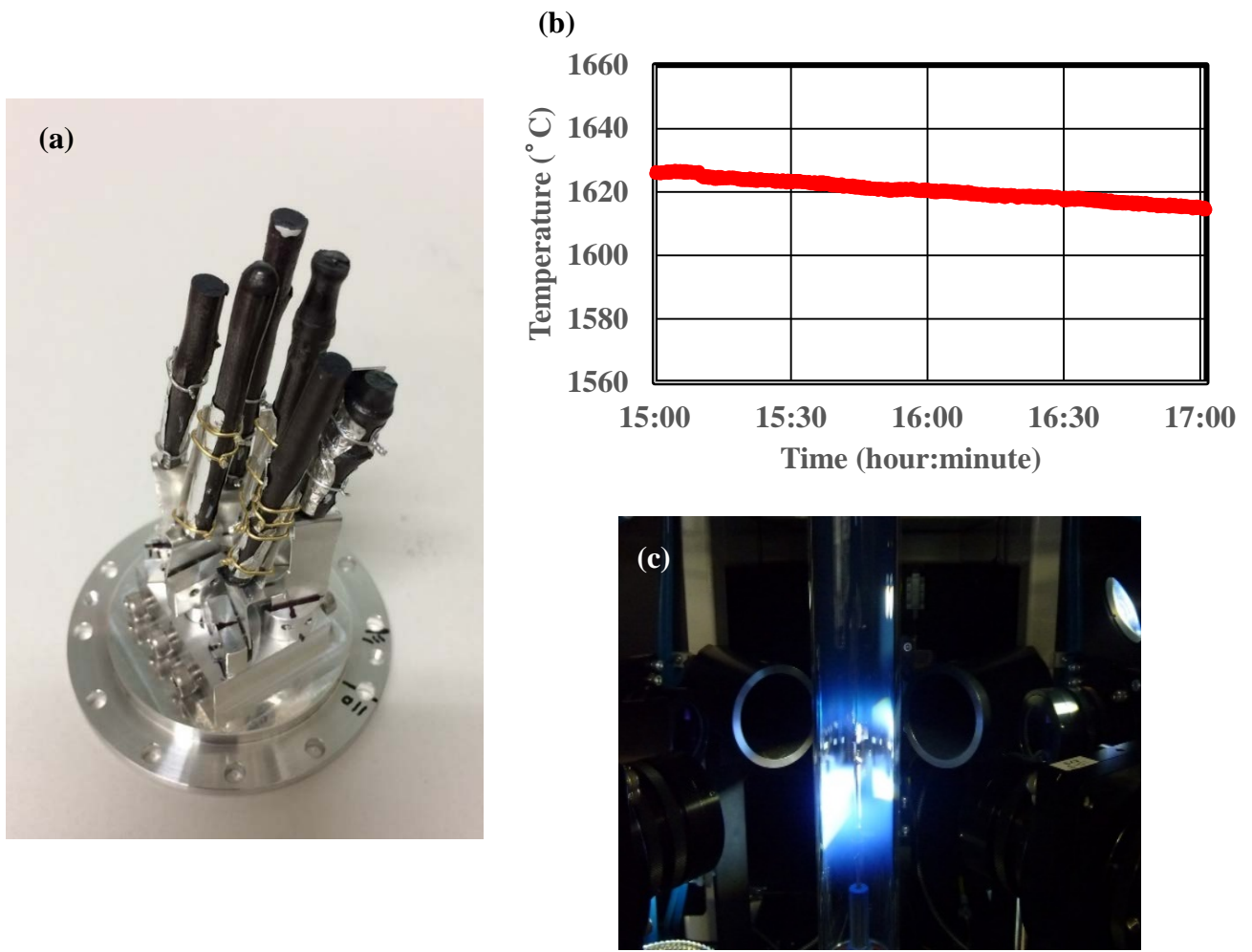

**Fig. 6. (a)** The photograph of single crystals of pyrochlore-type $Nd_2Mo_2O_7$ prepared for neutron scattering experiments, through the courtesy of Shinichi Ito, High Energy Accelerator Research Organization, Japan. All of the single-crystal rods have the crystal growing axis of around [1-10]-direction by using the seed crystal of [1-10]-direction. Each crystal axis of the rod has been accurately positioned at [1-10]-direction using X-ray diffraction instruments. In this case, LFZ is using type A with ±2mm flat shape in the vertical direction. **(b)** The elapsed-time dependence of the temperature at the molten zone of $Nd_2Mo_2O_7$. The molten zone temperature is observed to decrease gradually from 1630°C to 1620°C in several tens of minutes because the irradiation intensity slightly decreases due to the adhesion of evaporated materials on the inner surface of the quartz tube. And then it becomes constant after several hours. **(c)** The picture of a quartz tube after the crystal growth. The bluish deposit is $MoO_2$. The portion irradiated by the laser beam appears transparent because of the re-evaporation of the adhered materials.

p. 19

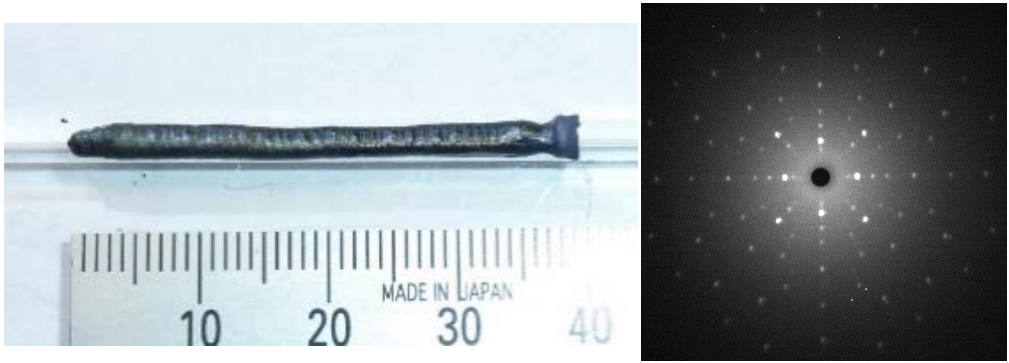

**Fig. 7.** The photograph (left panel) of the LFZ grown crystal of SrRuO$_3$. The clear Laue pattern from the cross-section of this rod (right panel) indicates that this rod is a single crystal. In this case, LFZ is using type B with ±10mm gaussian shape along the vertical direction.